\documentclass[prb,showpacs, twocolumn,amsmath,amssymb,superscriptaddress]{revtex4}
\usepackage{graphicx}% Include figure files
\usepackage{dcolumn}% Align table columns on decimal point
\usepackage{bm}% bold math
\pacs{74.43.-f}
\begin{document}

%\preprint{APS/123-QED}

\title{
Strain-induced enhancement of electric quadrupole splitting
in resistively detected nuclear magnetic resonance spectrum in quantum Hall systems
}

\author{M.~Kawamura}
\email{minoru@riken.jp}
\affiliation{Institute of Industrial Science,
 University of Tokyo, 4-6-1 Komaba, Meguro-ku, Tokyo 153-8505, Japan}
\affiliation{Advanced Science Institute, RIKEN, 2-1 Wako, Saitama 351-0198, Japan}
\affiliation{PRESTO, Japan Science and Technology Agency,
 4-1-8 Kawaguchi, Saitama 333-0012, Japan}
\author{T.~Yamashita}
\affiliation{Institute of Industrial Science,
 University of Tokyo, 4-6-1 Komaba, Meguro-ku, Tokyo 153-8505, Japan}
\author{H.~Takahashi}
\affiliation{Institute of Industrial Science,
 University of Tokyo, 4-6-1 Komaba, Meguro-ku, Tokyo 153-8505, Japan}
\author{S.~Masubuchi}
\affiliation{Institute of Industrial Science,
 University of Tokyo, 4-6-1 Komaba, Meguro-ku, Tokyo 153-8505, Japan}

\author{Y.~Hashimoto}
\affiliation{Institute for Solid State Physics, 
University of Tokyo, \\
5-1-5 Kashiwanoha, Kashiwa 277-8581, Japan}

\author{S.~Katsumoto}
\affiliation{Institute for Solid State Physics, 
University of Tokyo, \\
5-1-5 Kashiwanoha, Kashiwa 277-8581, Japan}
\affiliation{Institute for Nano Quantum Information Electronics, University of Tokyo,
4-6-1 Komaba, Meguro-ku, Tokyo 153-8505, Japan}
\author{T.~Machida}
	\email{tmachida@iis.u-tokyo.ac.jp}
\affiliation{Institute of Industrial Science,
 University of Tokyo, 4-6-1 Komaba, Meguro-ku, Tokyo 153-8505, Japan}
\affiliation{Institute for Nano Quantum Information Electronics, University of Tokyo,
4-6-1 Komaba, Meguro-ku, Tokyo 153-8505, Japan}

\date{\today}% It is always \today, today,

\begin{abstract}
We show electrical coherent manipulation of quadrupole-split nuclear spin states
in a GaAs/AlGaAs heterostructure
on the basis of the breakdown of quantum Hall effect.
The electric quadrupole splitting in nuclear spin energy levels is
intentionally enhanced by applying an external stress to the heterostructure.
Nuclear magnetic resonance spectra with clearly separated triple peaks are obtained,
and Rabi oscillations are observed between the nuclear spin energy levels.
The decay of the spin-echo signal is compared between the cases before
and after the enhancement of quadrupole splitting.
\end{abstract}

\maketitle

\begin{figure}[btp]
	\begin{center}
	\includegraphics[width=8cm]{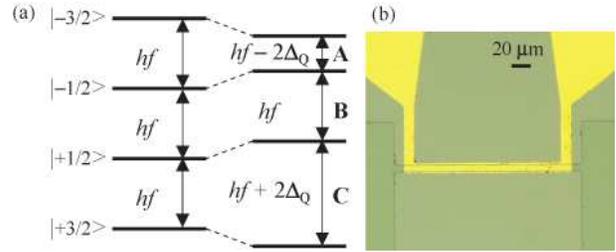}
	\caption{
		\label{schematic}
		(a) Energy diagram of $I$ = 3/2 nuclear spin system
			for $\Delta_{\rm Q}$ = 0 (left) and $\Delta_{\rm Q} \neq$ 0 (right).
		(b) Micrograph of the Hall-bar device.
			A pulsed rf-magnetic field is irradiated using the metal strip
			covering the 5-$\mu$m-wide conduction channel.
		}
	\end{center}
\end{figure}

Nuclear spins in semiconductors have recently attracted
considerable attention because
their extremely long coherence time is suitable for
the implementation of quantum bits/memories\cite{Kane1998, Ladd2002, Taylor2003}.
In order to manipulate nuclear spin quantum states coherently,
all-electrical\cite{Machida2003, Yusa2005, Takahashi2007}/optical
\cite{Salis2001, Sanada2006, Kondo2008} nuclear magnetic resonance (NMR)
techniques have been developed on the basis of the hyperfine interaction
between nuclear spins and electron spins.
However, all of  these new NMR techniques
 have been successfully applied in GaAs.
Since all the constituent atoms in GaAs have nuclear spins $I = 3/2$, 
the nuclear spin states split into four-level $I_z$ eigen states
$\vert \pm3/2\rangle$ and  $\vert \pm1/2\rangle$ 
in a magnetic field as shown on the left-hand side of Fig.~\ref{schematic}(a).
Such a four-level system can be regarded as coupled quantum bits
if transitions between any pairs of the levels
 are controlled selectively\cite{Leuenberger2003}.

In the presence of a local electric field gradient,
the electric quadrupole interaction produces
non-equidistant nuclear spin energy levels\cite{Slichter}
as shown on the right-hand side of Fig.~\ref{schematic}(a).
Although the electric quadrupole splitting energy $\Delta_{\rm Q}$
is zero in GaAs because of the cubic symmetry of the GaAs crystal, 
it is possible to increase the amplitude of $\Delta_{\rm Q}$
by applying an external stress to the crystal\cite{Guerrier, Knotz, Eickhoff},
because $\Delta_{\rm Q}$ is proportional to the local electric field gradient.
Such an external stress can be applied using pressure cells or
piezoelectric devices\cite{Guerrier, Knotz} or by coating the
surface of GaAs with different material\cite{Eickhoff}.

In this paper, we show intentional enhancement of electric quadrupole splitting
and selective control of a four-level nuclear spin system.
Using the breakdown phenomenon of quantum Hall effect,
nuclear spins are polarized and NMR are detected.
We apply an external stress
to a Hall-bar device by coating its surface with a polyimide film. 
NMR spectra with clearly separated triple peaks
are obtained in the polyimide-coated devices.
Splitting of NMR spectra enables us
to show the selective and coherent manipulation
of a four-level nuclear spin system using pulsed NMR techniques.
Furthermore, the decay of spin-echo signal is compared
between the cases before and after the enhancement of
quadrupole-splitting.

The experiments were performed using a two-dimensional electron gas (2DEG)
in a GaAs/Al$_{0.3}$Ga$_{0.7}$As single heterostructure wafer
grown by molecular beam epitaxy on a (001) oriented GaAs substrate.
The 2DEG is located at 230 nm below the surface.
The mobility and density of the 2DEG  at 4.2 K are 220~m$^2$/Vs and 
1.6~$\times$~10$^{15}$~m$^{-2}$, respectively.
Figure~\ref{schematic}(b) shows an optical micrograph of the Hall-bar device
used in the present study.
A 10-$\mu$m-wide Ti/Au Schottky gate electrode that was used for tuning electron density
also functioned as a local coil for generating radio-frequency (rf) magnetic fields $B_{\rm rf}$
parallel to the 2DEG.
External static magnetic field $B$ was applied %using a superconducting solenoid
perpendicular to the 2DEG, hence parallel to the [001] direction of the GaAs crystal.
All the measurements were performed at 50~mK using a $^3$He-$^4$He dilution refrigerator.
The sample chip (1 mm $\times$ 1 mm $\times$ 0.5 mm) was glued backside to a ceramic chip carrier
using silver paste. After wiring to the Hall-bar device,
the ceramic package was held against the cold finger plate of the dilution refrigerator
to make a good thermal contact.

\begin{figure}[tbp]
	\includegraphics[width=8.0cm]{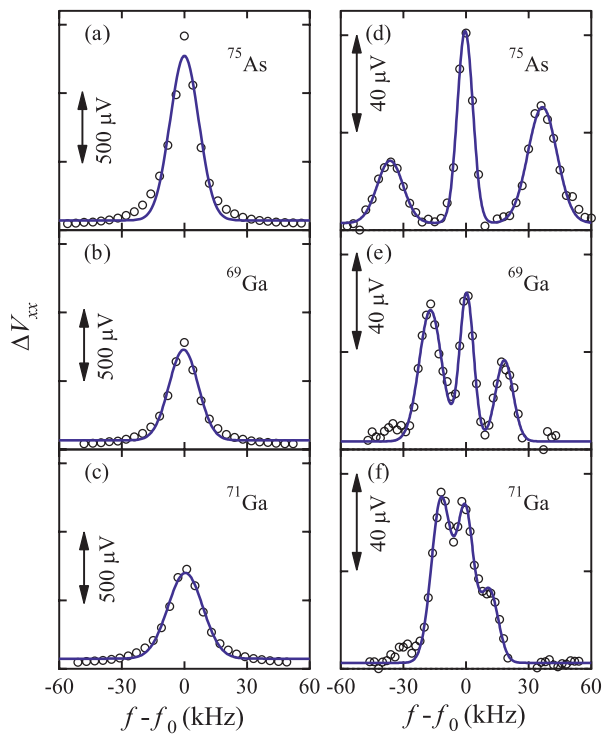}
	\caption{
		\label{spectrum}
		(a) - (c) NMR spectra obtained using the Hall-bar device before
		it was coated with a polyimide film.
		$B$ = 4.92 T ($\nu$ = 1.08). Values of $f_0$ are 35.520 MHz (a),
		 49.808 MHz (b), and  63.286 MHz (c).
		(d) - (f) NMR spectra obtained using the Hall-bar device after it was coated
		with the polyimide film.
		$B$ = 5.82 T ($\nu$ = 1.00). Values of $f_0$ are  41.963 MHz (d),
		 58.847 MHz (e), and  74.771 MHz (f).
		The solid curves are the fitting curves.
		}
\end{figure}

NMR signals were obtained by 
dynamic nuclear polarization (DNP) and resistive detection (RD) techniques
in a breakdown regime of integer quantum Hall effect (QHE),
as already demonstrated in our earlier studies\cite{Kawamura2007}.
As an initialization process, nuclear spins are dynamically polarized 
through the hyperfine interaction between nuclear spins and electron spins
under a breakdown regime of a quantum Hall state with the Landau level filling factor 
$\nu$ = 1.
By applying a bias current larger than the critical current of the QHE breakdown,
electrons are excited to the upper Landau subband, accompanied by flips of electron spins.
The flips of electron spins cause flops of nuclear spins via the hyperfine interaction,
resulting in positive nuclear polarization\cite{Kawamura2007} $\langle I_z \rangle >$ 0.
Then, the nuclear spin states are manipulated
by applying $B_{\rm rf}$.
The manipulated nuclear spin state is read out by measuring the longitudinal voltage $V_{xx}$
of the Hall-bar device.
The read-out procedure is based on the fact that the positively polarized nuclear spins
($\langle I_z \rangle > $0) reduce the Zeeman splitting energy of electrons,
which increases $V_{xx}$.

First, we measured the NMR spectrum using an uncoated Hall-bar device.
In Figs.~\ref{spectrum}(a), (b), and (c),
the changes in the longitudinal voltage $\Delta V_{xx}$ induced
by the $B_{\rm rf}$ irradiation are plotted as a function of the $B_{\rm rf}$ frequency.
Each curve corresponds to the NMR spectrum for $^{75}$As, $^{69}$Ga, and $^{71}$Ga.
A single-peak spectrum is observed for all the three nuclear species.
These single-peak spectra indicate that the nuclear spin levels
are distributed almost equidistantly
as illustrated in the left-hand panel in Fig.~\ref{schematic}(a).

Next, the Hall-bar device was warmed up to room temperature;
at this temperature, a droplet of polyimide solution
was dropped onto the surface of the device.
Then, the polyimide coating was baked in N$_2$ atmosphere at 180 $^\circ$C for 15 min.
The polyimide-coated device was cooled down again for the NMR measurements.\cite{carrierdensity}
Since the thermal shrinkage rate of the polyimide film is considerably higher than 
that of the GaAs film,
the subsequent cooling of the polyimide-coated device
is expected to induce a large strain in the device.

Figure~\ref{spectrum}(d) shows the NMR spectrum of $^{75}$As after
it was coated with the polyimide film.
The NMR spectrum is split into three peaks.
These peaks correspond to transitions A, B, and C
shown on the right-hand side of Fig.~\ref{schematic}(a).
The NMR spectra of $^{69}$Ga and $^{71}$Ga
are also split as shown in Figs.~\ref{spectrum}(e) and (f), respectively.
Amplitudes of the splitting are $\Delta f$ = 36 kHz, 18 kHz, and 11 kHz
for $^{75}$As, $^{69}$Ga, and $^{71}$Ga, respectively.
The ratio of $\Delta f$ is in good agreement with the ratio of the quadrupole moment $Q$:
$\Delta f({\rm ^{69}Ga})/\Delta f({\rm ^{71}Ga})$ = 1.6
agrees with $Q(^{69}{\rm Ga})/Q(^{71}{\rm Ga})$  
= $0.19\times 10^{-28}~{\rm m}^{2}/0.12\times 10^{-28}~{\rm m}^{2}$ = 1.6.
This indicates that the splitting of the NMR spectra 
is attributed to the electric quadrupole interaction.
We observed the splitting in the spectrum of $^{75}$As in another device;
the single-peak spectrum before the polyimide coating is split to three peaks ($\Delta f$ = 16 kHz)
after the polyimide coating.
The NMR peak splitting of 7.5 kHz for $^{75}$As was also observed
in yet another device after coating its surface
with PMMA electron-beam resist\cite{nosplit}.

\begin{figure}[tbp]
\includegraphics[width=8.0cm]{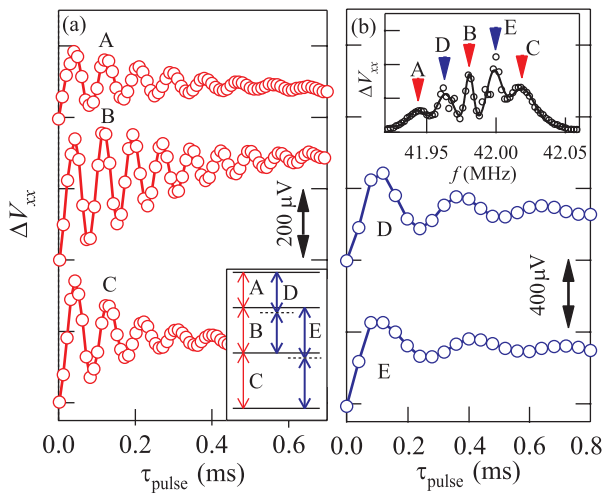}
	\caption{
		\label{rabis}
		Changes in $V_{xx}$ induced by applying a pulse of $B_{\rm rf}$
		with various pulse durations
		$\tau_{\rm pulse}$ at $B$ = 5.82 T ($\nu$ = 1.00).
		The frequencies of $B_{\rm rf}$ are
		 41.952 MHz (A),  41.987 MHz (B), and   42.022 MHz (C) in (a),
		and  41.969 MHz (D) and  42.007 MHz (E) in (b).
		The curves are offset for clarity.
		The inset in (a) shows schematic energy diagram 
		for single- and two-photon absorption/emission.
		The inset in (b) shows the NMR spectrum of $^{75}$As
		with the input rf-voltage $V_{\rm rf}$ = 4.8 V.
		}
\end{figure}

We consider that the polyimide film produces a strain
in the GaAs/AlGaAs heterostructure,
resulting in the generation of a large electric field gradient
at nuclear spin sites, and the induced electric field gradient
enhances $\Delta_{\rm Q}$.
Comparing the observed splitting of 36 kHz in the NMR spectrum of $^{75}$As [Fig. 2(d)]
with the earlier measurements in GaAs quantum wells\cite{Guerrier},
the strain in our device is estimated as 1.7 $\times$ 10$^{-4}$.
The estimated value of the strain seems consistent with 
the results of electron transport measurements of 2DEG
under a strain-induced periodic potential modulation\cite{Endo}.
From the FWHM of the NMR spectrum of $^{75}$As [Fig.~\ref{spectrum}(a)],
the strain in the device before the polyimide film coating 
is estimated to be not larger than 3.9 $\times$ 10$^{-5}$,
even if the broadening of the spectrum
is attributed to $\Delta_{\rm Q}$.
Therefore, contribution of the other sources of strain, 
such as Ti/Au Schotky gate or the silver paste on the backside
of the sample chip, is small compared to that of the polyimide film.

\begin{figure}[tbp]
	\includegraphics[width=7.0cm]{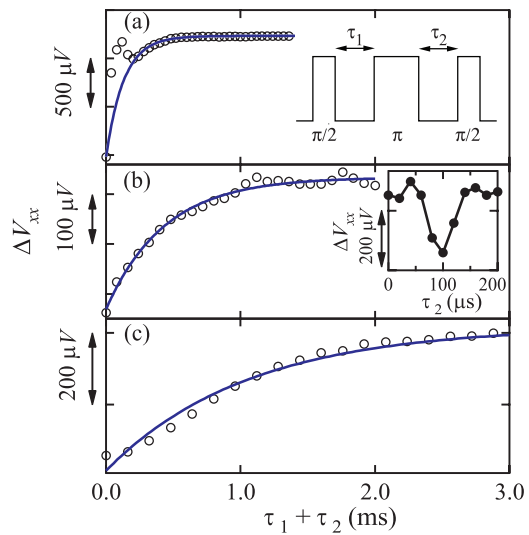}
	\caption{
		\label{echo}
		(a)	Decay of spin-echo signal obtained in the Hall-bar device
		before the polyimide coating. $B$ = 4.92 T ($\nu$ = 1.08).
		(b)-(c) Decays of spin-echo signals obtained in the Hall-bar device 
		after the polyimide coating. $B$ = 5.82 T ($\nu$ = 1.00).
		In the case of (c), the electrons were depleted
		during the rf pulse irradiation.
		The inset of (a) shows a schematic of the pulse sequence for the spin-echo measurements.
		The inset of (b) shows a representative spin-echo signal obtained by changing $\tau_2$
		with a fixed $\tau_1$ = 100 $\mu$s.
	}
\end{figure}

Figure~\ref{rabis}(a) shows the changes in $V_{xx}$ induced by applying a pulse 
of $B_{\rm rf}$ with various pulse durations $\tau_{\rm pulse}$ at $B$ = 5.82 T ($\nu$ = 1).
We note that the amplitude of $B_{\rm rf}$ in the pulsed NMR measurements
(Fig.~\ref{rabis}) is 12 times larger than that used 
to obtain the continuous-wave NMR spectra (Fig.~\ref{spectrum})\cite{power}.
The $B_{\rm rf}$ frequencies for the curves A, B, and C are
41.952 MHz, 41.987 MHz, and 42.022 MHz, respectively,
as indicated in the inset in Fig.~\ref{rabis}(b). 
The oscillatory changes in $\Delta V_{xx}$ denoted A, B, and C
correspond the Rabi oscillations of $^{75}$As
for transitions A ($\vert +3/2\rangle \leftrightarrow \vert +1/2\rangle$), 
B ($\vert +1/2\rangle \leftrightarrow \vert -1/2\rangle$),
and C ($\vert -1/2\rangle \leftrightarrow \vert -3/2\rangle$), respectively.
These results clearly show that the intentional enhancement of $\Delta_{\rm Q}$
enables the selective and coherent control of the four-level nuclear spin system.
Additional two peaks (D and E) are seen in the spectrum
at the middle frequencies between the peaks A and B, and B and C
as similar to the work by Yusa {\it et al}\cite{Yusa2005}.
These additional peaks correspond to the two-photon absorption/emission processes
($\vert +3/2\rangle \leftrightarrow \vert -1/2\rangle$ and
$\vert +1/2\rangle \leftrightarrow \vert -3/2\rangle$)
induced by the irradiation of $B_{\rm rf}$ with a large amplitude.
The oscillations D and E in Fig.~\ref{rabis}(b) correspond to the two-photon Rabi oscillations
taken at the $B_{\rm rf}$ frequencies of 41.969 MHz and 42.007 MHz, respectively.
The observed frequency of the two-photon Rabi oscillation $\Omega_{\Delta m = 2}$ = 3.8 kHz
nearly agrees with the calculated value\cite{Gentile}
$\Omega_{\Delta m = 2} \sim \Omega_{\Delta m = 1}^2/2\Delta_{\rm Q}$ 
=  (12.5 kHz)$^2$/36 kHz = 4.3 kHz.

We verify the effect of electric quadrupole splitting on the nuclear spin coherence
time by performing spin-echo experiments.
We applied a sequence of $\pi/2$-$\pi$-$\pi/2$ rf pulses\cite{Machida2004},
as shown in the inset of Fig.~\ref{echo}(a).
The inset of Fig.~\ref{echo}(b) shows a representative spin-echo signal 
in the device after the polyimide film coating
obtained by changing the second waiting time $\tau_2$ with a fixed 
first waiting time $\tau_1$ = 100 $\mu$s with the $B_{\rm rf}$ frequency of 42.022 MHz.
The coherence time $T_{2}$ is estimated from the decay of the spin-echo signal
by changing the total waiting time $\tau_1 + \tau_2$ under the condition $\tau_1 = \tau_2$.
Figure~\ref{echo}(a) shows 
the decay of the spin-echo signals for $^{75}$As
in the device before the polyimide film coating.
The $B_{\rm rf}$ frequency  was tuned to 41.963 MHz,
the peak frequency in Fig.~\ref{spectrum}(a),
where all the three NMR transitions occur simultaneously.
The value of $T_2$ is estimated to be no longer than 0.2 ms,
and the signal decays non-monotonically.
In contrast, after coating the Hall-bar device with the polyimide film,
the spin-echo signal decays exponentially as shown in Fig.~\ref{echo}(b).
The $B_{\rm rf}$ frequency was tuned to 42.022 MHz,
the peak C in the inset of Fig.~\ref{rabis}(b).
The value of $T_2$ is estimated as 0.42 ms, which is
almost twice longer than that obtained before the polyimide film coating.
The decay time of the Rabi oscillations 
is also increased after the polyimide film coating (not shown).
In addition, as shown in Fig.~\ref{echo}(c),
the value of $T_2$ is further increased to 1.1 ms
by decoupling the nuclear system from the electron system
during nuclear-spin manipulation\cite{Masubuchi2006, ota};
electrons are depleted by applying negative dc voltage to the Schottky gate electrode
during the rf-pulse irradiation.

In summary, we have demonstrated strain-induced enhancement
of the electric quadrupole splitting and electrical coherent manipulation
in $I$ = 3/2 nuclear spin energy levels 
in GaAs/GaAs heterostructure.
The DNP and RD techniques used in the present study
can be employed at temperatures higher than 1 K
and even in a 2DEG with a relatively low electron mobility\cite{Takahashi2007,Kawamura2007},
because the techniques are based on the breakdown phenomena of QHE.

%\begin{acknowledgments}
This work was supported by a Grant-in-Aid from 
MEXT, the Sumitomo Foundation, 
and the Special Coordination Funds for Promoting 
Science and Technology.
%\end{acknowledgments}

\newpage


\begin{thebibliography}{99}
\bibitem{carrierdensity}
	The carrier density of the 2DEG was increased by approximately 10\%
	after the polyimide coating.
\bibitem{nosplit}
	We measured other uncoated Hall-bar devices (more than 10)
	fabricated from the same wafer,
	and the quadrupole-splitting was not resolved in these uncoated devices.
\end{thebibliography}
\end{document}